\documentclass[11pt]{article}

\usepackage{amssymb}
\usepackage{amsmath,amssymb}
\usepackage{amsthm}
\usepackage[singlelinecheck=off]{caption}
\usepackage{microtype}
\usepackage{graphicx}
\usepackage{algorithm,algpseudocode}
\usepackage{mathrsfs}
\usepackage{epstopdf}
\usepackage{float}
\usepackage{url}
\newcommand{\Tr}{\mathrm{Tr}}

\newcommand{\Sgt}{S}
\newcommand{\sgt}{s}

\newtheorem{observation}{Observation}

\title{Bias Correction in Saupe Tensor Estimation}

\author{
Y. Khoo\thanks{Department of Physics, Princeton University, Princeton, NJ 08544,USA ({\tt ykhoo@princeton.edu}).}
\and
A. Singer\thanks{Department of Mathematics and PACM, Princeton University, Princeton, NJ 08544, USA ({\tt amits@math.princeton.edu}).}
\and
D. Cowburn\thanks{Department of Biochemistry, Albert Einstein College of Medicine, Bronx, NY 10461, USA ({\tt david.cowburn@einstein.yu.edu}).}
}
\begin{document}
\maketitle
\begin{abstract}
Estimation of the Saupe tensor is central to the determination of molecular structures from residual dipolar couplings (RDC) or chemical shift anisotropies. Assuming a given template structure, the singular value decomposition (SVD) method proposed in \cite{losonczi1999order} has been used traditionally to estimate the Saupe tensor. Despite its simplicity, whenever the template structure has large structural noise, the eigenvalues of the estimated tensor have a magnitude systematically smaller than their actual values. This leads to systematic error when calculating the eigenvalue dependent parameters,  magnitude and rhombicity. We propose here a Monte Carlo simulation method to remove such bias. We further demonstrate the effectiveness of our method in the setting when the eigenvalue estimates from multiple template protein fragments are available and their average is used as an improved eigenvalue estimator. For both synthetic and experimental RDC datasets of ubiquitin, when using template fragments corrupted by large  noise, the magnitude of our proposed bias-reduced estimator generally reaches at least 90\% of the actual value, whereas the magnitude of SVD estimator can be shrunk below 80\% of the true value.
\end{abstract}

\section{Introduction}

\subsection{Background}
The residual dipolar couplings (RDC) of molecules can be measured when the molecule ensemble in solution exhibits partial alignment with the magnetic field in a nuclear magnetic resonance (NMR) experiment. Due to the $r^{-3}$ dependence, such effects can be measured with high precision and provides accurate alignment information of a specific single or multiple bond vector to the magnetic field. This is in contrast to the Nuclear Overhauser Effect (NOE) which scales as  $r^{-6}$. Hence the importance of RDC in obtaining high quality protein structures and studying molecular dynamics has increased considerably over the last decade. For a detailed survey of RDC and its applications we refer readers to \cite{lipsitz2004rdc,blackledge2005survey,tolman2006rdc,salmon20154539}.
% also add   {Salmon, 2015 #4539} 1.	Salmon L, Blackledge M. Investigating protein conformational energy landscapes and atomic resolution dynamics from nmr dipolar couplings: A review. Reports on Progress in Physics. 2015;78 (12):126601 https://drive.google.com/open?id=0B8ZTrC2gf5lKZDFra0RLWFZhWEE

We first give a brief summary of RDC. Let $v_{nm}$ be the unit vector denoting the direction of the vector between nuclei $n$ and $m$. Let $b$ be the unit vector denoting the direction of the magnetic field. The RDC $D_{nm}$ due to the interaction between nuclei $n$ and $m$ is
\begin{equation}
D_{nm} = D_{nm}^{max} \left \langle \frac{3 (b^T v_{nm})^2 -1}{2} \right \rangle_{t,e}.
\end{equation}
$D_{nm}^{max}$ is a constant depending on the gyromagnetic ratios $\gamma_n,\gamma_m$ of the two nuclei,  bond length $r_{nm}$, and  Planck's constant $h$ as
\begin{equation}
D_{nm}^{max} = -\frac{\gamma_n \gamma_m h}{2 \pi^2 r_{nm}^3},
\end{equation}
and $\langle\ \cdot \rangle_{t,e}$ denotes the ensemble and time averaging operator. As presented, RDC depends on the relative angle between the magnetic field and the bond. Extracting such angular information from RDC  complements NOE and  other measurements for determining the molecular structure.
%\today include REF {Vögeli, 2015 #4540} 1.	Vögeli B, Olsson S, Riek R, Güntert P. Complementarity and congruence between exact noes and traditional nmr probes for spatial decoding of protein dynamics. Journal of structural biology. 2015;191 (3):306-17 https://drive.google.com/open?id=0B8ZTrC2gf5lKTHc4NmtGU3RPcjA
%provide the orientational information for two bonds that are far apart, unlike the case of NOE.
%It is conventional to interpret the RDC measurement in the molecular frame. More precisely, we treat the molecule as being static in some coordinate
It is conventional to interpret the RDC measurement in the molecular frame. More precisely here, for this analysis of bias, we treat the molecule as being static in some coordinate system, and the magnetic field direction being a time and sample varying vector. In this case the RDC becomes
\begin{equation}
\label{RDC}
D_{nm} = D_{nm}^{max} v_{nm}^T S v_{nm},
\end{equation}
where the Saupe tensor $S$ is defined as
where the Saupe tensor $S$ \cite{Saupe1964} is defined as
\begin{equation}
\label{Saupe tensor}
S = \frac{1}{2}(3 B - I_3),\qquad B =  \left \langle bb^T \right \rangle_{t,e}.
\end{equation}
%$B$ is known as the field tensor and $I_3$ denotes the $3\times 3$ identity matrix. We note that $S$ is symmetric and $\Tr(S) = 0$. In order to use RDC for structural refinement of a protein, $S$ is usually first determined from a proposed structure (known $v_{nm}$) that is similar to the protein under study. To satisfy the assumption that the molecule is static in the molecular frame, a rigid fragment of the known structure has to be selected. $S$ can be determined if the fragment contains sufficiently many RDC measurements.
$B$ is known as the field tensor and $I_3$ denotes the $3\times 3$ identity matrix. We note that $S$ is symmetric and $\Tr(S) = 0$. In order to use RDC for structural refinement of a protein, $S$ is usually either first determined from a proposed structure (known $v_{nm}$) that is similar to the protein under study, or estimated from a distribution of RDCs \cite{clore19989654491} . To satisfy the assumption that the molecule is static in the molecular frame, a rigid fragment of the known structure has to be selected. $S$ can be determined if the fragment contains sufficient RDC measurements.
%{Clore, 1998, 9654491}  	Clore GM, Gronenborn AM, Bax A. A robust method for determining the magnitude of the fully asymmetric alignment tensor of oriented macromolecules in the absence of structural information. Journal of magnetic resonance. 1998;133(1):216-21  10.1006/jmre.1998.1419

\subsection{Notation}
We summarize here the notation that is used throughout the paper.
% Our notation is consistent with existing literature on RDC and Saupe tensor estimation.
For a $3\times 3$ matrix $A$, we use $A_{ij}$, $i,j = x,y,z$ to denote the nine entries of the matrix. When $A$ is symmetric, we denote the eigen-decomposition of $A$ by
$$ A = U(A) \Lambda(A) U(A)^T,$$
where $U(A)$ is an orthogonal matrix (i.e. $U(A)^T U(A) = U(A) U(A)^T = I_3$) and $\Lambda(A)$ is a diagonal matrix
\begin{equation}
\begin{bmatrix} \lambda_x(A) & 0 & 0 \\ 0 & \lambda_y(A) & 0 \\ 0 & 0 & \lambda_z(A) \end{bmatrix}
\end{equation}
that contains the eigenvalues of $A$ on the diagonal in ascending order. For a matrix $A\in \mathbb{R}^{n\times n}$ we use
$$\|A\|_F = \sqrt{\sum_{i,j=1}^n A_{ij}^2}$$
%to denote the Frobenius norm of the matrix. For a vector $v$, we often use $v_i$ to denote its $i$-th entry, and $i=1,\ldots,n$ if $v\in \mathbb{R}^n$. In the special case of $v\in \mathbb{R}^3$, we use $v_x,v_y,v_z$ to denote each entry of the vector $v$. For a matrix $A$, we use $A_i$ to denote its $i$-th column.
to denote the Frobenius norm of the matrix. For a vector $v$,  $v_i$  denotes its $i$-th entry, and $i=1,\ldots,n$ if $v\in \mathbb{R}^n$. In the special case of $v\in \mathbb{R}^3$, we use $v_x,v_y,v_z$ to denote components of the vector $v$. For a matrix $A$, we use $A_i$ to denote its $i$-th column.
%\begin{equation}
%\begin{bmatrix} \lambda_x(A) & 0 & 0 \\ 0 & \lambda_y(A) & 0 \\ 0 & 0 & \lambda_z(A) \end{bmatrix}
%\end{equation}
 %For a vector $v$, we often use $v_i$ to denote its $i$-th entry, and $i=1,\ldots,n$ if $v\in \mathbb{R}^n$. In the special case of $v\in \mathbb{R}^3$, we use $v_x,v_y,v_z$ to denote each entry of the vector $v$.  We use $I_n$ to denote the $n\times n$ identity matrix. We use the pair $(n,m)\in E_\text{RDC}$ to index the bond between nuclei $n$ and $m$ and $E_\text{RDC}$ to denote the set consisted of all bonds with RDC being measured. The number of RDC measurements is denoted by $M$, where $M = \vert E_\text{RDC} \vert$. We denote estimators by hat accent, for example $\hat \theta$ stands for the estimate for the underlying parameter $\theta$.

\subsection{Previous approach}
We review the singular value decomposition (SVD) approach \cite{losonczi1999order} for estimating the Saupe tensor.
%Using the fact that $S$ is symmetric and $\Tr(S) = 0$, eq. (\ref{RDC}) can be rewritten as
 $S$ is symmetric and $\Tr(S) = 0$, so eqn. (\ref{RDC}) can be rewritten as
\begin{multline}
\label{RDC for bond nm}
D_{nm}/D_{nm}^{max} = ({v_{nm}}_y^2-{v_{nm}}_x^2) S_{yy} + ({v_{nm}}_z^2-{v_{nm}}_x^2) S_{zz} \cr
%+ 2 {v_{nm}}_x {v_{nm}}_y S_{xy} + 2 {v_{nm}}_x {v_{nm}}_z S_{xz} + 2 {v_{nm}}_y {v_{nm}}_z S_{yz},
+ 2 {v_{nm}}_x {v_{nm}}_y S_{xy} + 2 {v_{nm}}_x {v_{nm}}_z S_{xz} + 2 {v_{nm}}_y {v_{nm}}_z S_{yz}
\end{multline}
%where ${v_{nm}}_i$, $i=x,y,z$ are the different components of $v_{nm}$ in the molecular frame. Hereafter we let $d_{nm} = D_{nm}/D_{nm}^{max}$, to which we refer as the RDC measurements. When there are $M$ RDC measurements, eq. (\ref{RDC for bond nm}) results in $M$ linear equations in five unknowns ($S_{yy},S_{zz},S_{xy},S_{xz}$ and $S_{yz}$), that can be written in matrix form as
where ${v_{nm}}_i$, $i=x,y,z$ are the different components of $v_{nm}$ in the molecular frame. We let $d_{nm} = D_{nm}/D_{nm}^{max}$, which are  the RDC measurements. When there are $M$ RDC measurements, eq. (\ref{RDC for bond nm}) results in $M$ linear equations in five unknowns ($S_{yy},S_{zz},S_{xy},S_{xz}$ and $S_{yz}$), that can be written in matrix form as
\begin{equation}
\label{ls prestegard}
A s = d,\qquad s =  \begin{bmatrix} S_{yy}\\S_{zz}\\S_{xy}\\S_{xz}\\S_{yz}  \end{bmatrix} \in \mathbb{R}^5, \qquad d =\begin{bmatrix} d_{n_1m_1}\\ \vdots \\d_{n_M m_M} \end{bmatrix} \in \mathbb{R}^M,
\end{equation}
and $A\in \mathbb{R}^{M\times 5}$.

Let the SVD of matrix $A$ be
\begin{equation}
A = U\Sigma V^T,
\end{equation}
where $U \in \mathbb{R}^{M\times 5}$ is a column orthogonal matrix (i.e. $U^T U = I_5$), $V\in \mathbb{R}^{5\times 5}$ is an orthogonal matrix, and $\Sigma \in \mathbb{R}^{5\times 5}$ is a positive diagonal matrix. We assume that $M\geq5$ and that $A$ has full rank for otherwise there is no unique solution to the linear system (\ref{ls prestegard}). The estimator of the Saupe tensor entries $s$ proposed in \cite{losonczi1999order} is
\begin{equation}
\label{SVD method}
\hat s = V \Sigma^{-1} U^T d.
\end{equation}
This is equivalent to the ordinary least squares (OLS) solution to the linear system (\ref{ls prestegard}), given by
\begin{equation}
\label{OLS estimator}
\hat s = (A^T A)^{-1} A^T d.
\end{equation}
For this reason, we will refer to this SVD method for Saupe tensor estimation as the OLS method. The computational aspects of employing the expressions in (\ref{SVD method}) and (\ref{OLS estimator}) are discussed in \cite{wirz2015fitting}. Notice that the Saupe tensor estimator given by (\ref{SVD method}) and (\ref{OLS estimator}), denoted $\hat S$, is the solution to the optimization problem
\begin{equation}
%\underset{S}{\min} \sum_{i=1}^M \vert d_{n_im_i} -  v_{n_im_i}^T S v_{n_im_i} \vert^2 \quad s.t.\ S\  \text{is symmetric},\ \Tr(S) = 0.
\underset{S}{\min} \sum_{i=1}^M \vert d_{n_im_i} -  v_{n_im_i}^T S v_{n_im_i} \vert^2 \quad \text{such that} \ S\  \text{is symmetric},\ \Tr(S) = 0.
\end{equation}
As such, the OLS estimator is also the maximum likelihood estimator when the error on $d_{nm}$ is assumed to be white Gaussian noise.

%Remark: Although the SVD procedure ensures $\Tr(S) = 0$ and $S$ being symmetric, it does not ensure that $S$ must be derived from a specific linear combination of a field tensor and the identity matrix as expressed in Eq. (\ref{Saupe tensor}). Instead, we can use a positive semidefinite matrix description of the field tensor $B$ to exactly characterize $S$. A matrix is positive semidefinite (PSD) if it is symmetric and has nonnegative eigenvalues. The field tensor $B$ can be characterized by the following observation:  NOT SURE WHY THIS SHOULD BE UNDER PLAYED ...
Although the SVD procedure ensures $\Tr(S) = 0$ and $S$  symmetric, it does not ensure that $S$ is obligatorily  derived from a specific linear combination of the field tensor and the identity matrix as expressed in eqn. (\ref{Saupe tensor}). In contrast, we can use a positive semidefinite matrix description of the field tensor $B$ to exactly characterize $S$. A matrix is positive semidefinite (PSD) if it is symmetric and has nonnegative eigenvalues. The field tensor $B$ can be characterized by the following observation:
\begin{observation}
\label{observtion:sdp}
$B = \langle b b^T \rangle_{t,e}$ where $b$ is a unit vector $\Leftrightarrow$ $B$ is positive semidefinite and $\Tr(B) = 1$.
\end{observation}
This follows from the convexity of both the set of PSD matrices and the set of unit trace matrices. A set is convex if and only if any weighted average of the elements in the set belongs to the set. Since for any unit vector $b$, $bb^T$ is PSD and $\Tr(bb^T) = \Tr(b^T b) = 1$, the time and ensemble average of such matrices is PSD with unit trace. Using this observation, the set of physical Saupe tensors can be characterized as
\begin{equation}
\mathcal{S} = \left\{S = \frac{1}{2}(3 B - I_3)\,\vert\, B\  \text{is PSD}, \Tr(B) = 1\right\}.
\end{equation}
However, since $B= (1/3)(2 S + I_3)$, when Saupe tensor $S$ has small entries it is dominated by $I_3$ and $B$ is always positive semidefinite. This is often the case in practice, hence the SVD method suffices for Saupe tensor estimation at this approximation. We note that should we ever need to estimate $S$ with large entries of magnitude around $10^{-1}$, we can solve
\begin{eqnarray}
\label{Saupe SDP}
&\underset{S\in\mathcal{S}}{\min}& \sum_{i=1}^M \vert d_{n_im_i} -  v_{n_im_i}^T S v_{n_im_i} \vert^2,
\end{eqnarray}
using semidefinite programming toolboxes available, e.g., in \texttt{CVX} \cite{cvx} so that the derived Saupe tensor remains physically reasonable.

\subsection{An Alternative Approach}
Here, we illustrate how the noise on the bond vectors $v_{nm}$ leads to bias in the OLS estimation of Saupe tensor parameters. We call such type of noise structural noise.
%By structural noise, we mean the template structure used for Saupe tensor fitting differs from the true structure of the molecule due to the flexibility of the protein.
In particular, we consider the situation when using noisy template structures differing from the true structure of the molecule for Saupe tensor fitting.
%Our simulation shows that when noise is added to the torsion angles of the protein backbone, the magnitude of the estimated Saupe tensor eigenvalues are typically smaller than their ground truth value, as demonstrated in Fig.~\ref{fig:RMS_vs_sigma}.
This situation may arise when using homologous structure in Saupe tensor fitting, or when the protein of interest has small conformation changes due to the dynamic nature of protein in solution.
Our simulation consisted of adding modest noise to the backbone torsion angles of the protein backbone.
This results in the magnitude of the estimated Saupe tensor eigenvalues being typically smaller than their true value, as demonstrated in Fig.~\ref{fig:RMS_vs_sigma}.
Our observation corroborates with the simulation results reported in \cite{zweckstetter2002evaluation}, in which independently and identically distributed (i.i.d.) random noise \cite{wasserman2013all}  is added to each bond vector instead.
% {Cover, 1969 #4582}  	Cover TM. Hypothesis Testing with Finite Statistics. The Annals of Mathematical Statistics. 1969;40(3):828-35
In linear regression, such decrease in magnitude of the estimator in the presence of noise on the regressor is commonly known as \emph{attenuation} \cite{carroll2006measurement}.
%While the focus of \cite{zweckstetter2002evaluation} is mainly to use Monte Carlo simulation to evaluate the uncertainty of estimated alignment magnitude and rhombicity, we focus on using it to correct the attenuation effect in the OLS Saupe tensor eigenvalues estimator.
While the focus of \cite{zweckstetter2002evaluation} is mainly to use Monte Carlo simulation to evaluate the uncertainty of estimated alignment magnitude and rhombicity, we here focus on using it to correct the attenuation effect in the OLS Saupe tensor eigenvalues estimator.
The method we propose bears similarity with the statistical method \emph{simulation extrapolation} (SIMEX) \cite{cook1994simex,stefanski1995simex} that is frequently used to correct for the attenuation effect.
Typically this type of methods are parametric and require noise variance as input. We show that an estimator of the noise magnitude can be obtained from the root mean square (RMS) of the residual of OLS estimator. We further demonstrate the usefulness of removing such bias when estimating the Saupe tensor eigenvalue from homology fragments of ubiquitin, using RDCs measured in two different alignment medias.
%We note that there are other approaches to improve the estimation of Saupe tensor in the presence of structural noise by studying local bond orientations using multiple alignment medias \cite{meiler2001mfa,tolman2002didc,meirovitch2012standard,sabo2014orium}.
We note that there are other approaches to improve the estimation of Saupe tensor in the presence of structural noise by studying local bond orientations using multiple alignment media \cite{meiler2001mfa,tolman2002didc,meirovitch2012standard,sabo2014orium}.
However here, we seek to remove the bias in the Saupe tensor eigenvalues in a single alignment media, when multiple Saupe tensor estimates is available from a collection of predetermined molecular fragments.

\section{Method}
\label{section:method}
%We now introduce a Monte Carlo method for correcting the bias in the eigenvalues of the OLS estimator arising from structural noise. note that rdc can be measured elsewhere than the backbone
We now introduce a Monte Carlo method for correcting the bias in the eigenvalues of the OLS estimator arising from structural noise of backbone torsion angles.
%For a protein with $N+1$ peptide planes, we assume the $\{\phi_i,\psi_i\}_{i=1}^N$ torsion angles fully determine the backbone conformation.
For a protein with $N+1$ peptide planes, we assume the $\{\phi_i,\psi_i\}_{i=1}^N$ torsion angles fully determine the backbone conformation, i.e. variations of $\{\omega_i\}_{i=1}^N$ are minimal.
The template structure's torsion angles $\phi_i^t, \psi_i^t$'s are related to the true structure via
\begin{equation}
\label{torsion angle noise}
%\phi_i^t = \phi_i + \sigma \alpha_i,\qquad \psi_i^t = \psi_i + \sigma \beta_i,\quad i=1,\ldots,N,
\phi_i^t = \phi_i + \sigma \alpha_i,\qquad \psi_i^t = \psi_i + \sigma \beta_i,\quad i=1,\ldots,N
\end{equation}
where $\alpha_i, \beta_i$'s are i.i.d. random normal variables with mean 0 and variance 1, and $\sigma$ is the level of noise on the torsion angles. Henceforth for a variable $\theta$, we make explicit the dependence on the torsion angles and noise by writing $\theta$ as $\theta(\phi_i^t,\psi_i^t)$. We also assume that the normalized dipolar coupling $d$ is noiseless, i.e. $d = A(\phi_i,\psi_i) s$, where $s$ corresponds to the entries of the "ground truth" Saupe tensor $S$. The validity of this assumption is discussed below in section \ref{section:add noise}. Our method consists of the following steps:

(1) Compute
$$\hat s(\phi_i^t,\psi_i^t) = (A(\phi_i^t,\psi_i^t)^T A(\phi_i^t,\psi_i^t))^{-1}  A(\phi_i^t,\psi_i^t)^T d$$.

(2) Generate $n_1$ copies of $A_{\text{sim}} = A(\phi_i^t + \sigma \alpha_i, \psi_i^t + \sigma \beta_i)$ by adding i.i.d. Gaussian noise with variance $\sigma^2$ to the torsion angles of the template structure.

(3) Find $$\hat s_{\text{sim}} = \hat s(\phi_i^t + \sigma \alpha_i, \psi_i^t + \sigma \beta_i) = (A^T_{\text{sim}} A_{\text{sim}})^{-1} A_{\text{sim}}^T d.$$

(4) Let $\hat S$ and $\hat S_{\text{sim}}$ be the Saupe tensor estimators corresponding to $\hat s$ and $\hat s_{\text{sim}}$. Let
$$\widehat{\text{Bias}} = \langle \Lambda(\hat S_{\text{sim}}) \rangle_{\text{sim}} - \Lambda(\hat S)$$
denote the bias estimate for the eigenvalues of the OLS estimator $\hat S$, where $\langle \cdot \rangle_{\text{sim}}$ denotes the averaging over $n_1$ simulated template structures.
We propose deriving $\Lambda$ (Eqn 5) from
$$\tilde \Lambda = \Lambda(\hat S) - \widehat{\text{Bias}} = 2 \Lambda(\hat S) - \langle \Lambda(\hat S_{\text{sim}}) \rangle_{\text{sim}}$$
as an estimator with less bias.

%The rationale of our method relies on the intuition that upon adding noise of similar magnitude to the linear system (\ref{ls prestegard}), the eigenvalues of the OLS estimator for the simulated samples should be biased away from $\Lambda(\hat S)$ by an amount similar to difference between $\Lambda(\hat S)$ and the ground truth $\Lambda(S)$.
The rationale  relies on the notion that upon adding noise of similar magnitude to the linear system (\ref{ls prestegard}), the eigenvalues of the OLS estimator for the simulated samples should be biased away from $\Lambda(\hat S)$ by an amount similar to the difference between $\Lambda(\hat S)$ and the true  $\Lambda(S)$
This is also the intuition behind
twicing \cite{RN5009,RN5006}, and related bootstrapped \cite{efron1994bootstrap} biased reduced estimators. Alternatively, one can understand this procedure from the viewpoint of the SIMEX technique \cite{cook1994simex} for correcting bias resulting from regressor noise. Under the SIMEX estimation framework one would simulate $A_{\text{sim}} = A(\phi_i^t + k\sigma \alpha_i, \psi_i^t + k\sigma \beta_i)$ with noise magnitudes of $k \sigma$ for various positive $k$ to find out the dependency of $\Lambda(\hat S_{\text{sim}})$ on $k$. The $k=0$ point corresponds to the case when no additional simulated noise is added, i.e. when the eigenvalue estimator is $\Lambda(\hat S)$. From the extrapolation of the relation between $\Lambda(\hat S_{\text{sim}})$ and $k$ one can obtain a debiased estimator at $k=-1$. Our method corresponds to the special case of SIMEX where we only add simulated noise with magnitude $k \sigma$ where $k=1$. Our numerical results shows that this suffices for the application of Saupe tensor eigenvalue estimation.

\subsection{Estimating $\sigma$}
\label{section:sigma estimate}
%We note that there is a caveat when using this parametric Monte Carlo method, in that it requires knowledge of the noise magnitude $\sigma$. Let the residual of the OLS estimator be defined as
We note that there is a general caveat when using any parametric Monte Carlo method, in that it requires knowledge of the noise magnitude $\sigma$. Let the residual of the OLS estimator be defined as
$$r \equiv d - A \hat s.$$
In the simple case when additive noise with variance $\sigma_{\text{add}}^2$ is added to the normalized dipolar couplings $d$, and $A$ has no structural noise, i.e. $A = A(\phi_i,\psi_i)$, the dependence between the RMS of the residual, denoted $\text{RMS}(r)$ and the noise magnitude can be readily calculated. In particular, an unbiased estimator of $\sigma_\text{add}^2$ is given by \cite{gross2003linear}
$$  \widehat{ \sigma_{\text{add}}^2} =  \frac{M}{M-5} \text{RMS}(r)^2. $$ %same as small m used previously? Define M ??
where $M$ is the number of linear equations.
%$$  \widehat{ \sigma_{\text{add}}^2} =  \frac{m}{m-5} \text{RMS}(r)^2. $$

Now in the case when there is noise on the design matrix $A = A(\phi_i^t,\psi_i^t)$ due to noise on the torsion angles (\ref{torsion angle noise}), we show that there exists a linear dependence of $\text{RMS}(r)$ on $\sigma$. We define $A_0 = A(\phi_i,\psi_i)$, and $A(\phi_i^t,\psi_i^t) = A_0 + E$. In this notation, normalized RDC $d = A_0 s$. Then
\begin{eqnarray}
\|r\|^2_2 &=& \|d - A \hat s \|_2^2\cr
&=&\|A_0 s - A (A^T A)^{-1} A^T (A_0 s)\|_2^2\cr
&=&s^T A_0^T (I_M - A (A^T A)^{-1} A^T) A_0 s.
\end{eqnarray}
The second equality follows from the fact that $I_M - A (A^T A)^{-1} A^T$ is a projection matrix. From
\begin{eqnarray}
&\ &A_0^T (I_M - A (A^T A)^{-1} A^T) A_0\cr
&=& A_0^T A_0 - (A-E)^T A (A^T A)^{-1} A^T (A-E) \cr
&=&A_0^T A_0 - A^T A + E^T A + A^T E - E^T A (A^T A)^{-1} A^T E\cr
&=&A_0^T A_0 - (A-E)^T (A-E) +E^T E - E^T A (A^T A)^{-1} A^T E\cr
&=&E^T (I_M-A (A^T A)^{-1} A^T) E,
\end{eqnarray}
we get
\begin{eqnarray}
\label{RMS residual}
 \|r\|^2_2 &=& s^T E^T (I_M-A (A^T A)^{-1} A^T) E s \cr
&\approx&  s^T E^T (I_M-A_0 (A_0^T A_0)^{-1} A_0^T) E s  \cr
&=&  s^T E^T P E  s
\end{eqnarray}
where $P = I_M-A_0 (A_0^T A_0)^{-1} A_0^T$ is a projection operator projecting vectors in $\mathbb{R}^M$ to $\mathbb{R}^{M-5}$. We drop the terms involving entries of $E$ raised to the power greater than 2 to obtain the approximation in (\ref{RMS residual}). Using Taylor expansion,
\begin{eqnarray}
E_{ij} &=& A_{ij} - {A_0}_{ij}\cr
&\approx& \sum_{k=1}^N \frac{\partial A_{ij}(\phi_k^t,\psi_k^t)}{\partial \phi_k^t}\bigg\vert_{\phi_k,\psi_k} \sigma \alpha_k + \frac{\partial A_{ij}(\phi_k^t,\psi_k^t)}{\partial \psi_k^t}\bigg\vert_{\phi_k,\psi_k} \sigma \beta_k\cr
&=& F_{ij} \sigma.
\end{eqnarray}
Plugging this into (\ref{RMS residual}), it is clear that $\|r\|^2_2$ depends linearly on $\sigma^2$ and
\begin{equation}
\label{Average RMS}
\langle \text{RMS}(r)^2 \rangle_{\alpha_i,\beta_i}\approx \frac{1}{M}\langle s^T F^T P F s \rangle_{\alpha_i,\beta_i} \sigma^2
\end{equation}
in the small noise regime. We therefore use
$$\hat \sigma = \sqrt{\frac{M}{ s^T F^T P F s}} \text{RMS}(r)$$
as the approximate noise magnitude when using the Monte Carlo method for bias reduction. Although we do not have the parameters $s, F$ and $P$ derived from the ground truth Saupe tensor and conformations, we can use $\hat s$ as surrogate of $s$, and use the noisy structure to derive an approximation of $F$ and $P$.

\begin{figure}
\centering
\includegraphics[width=0.9\textwidth]{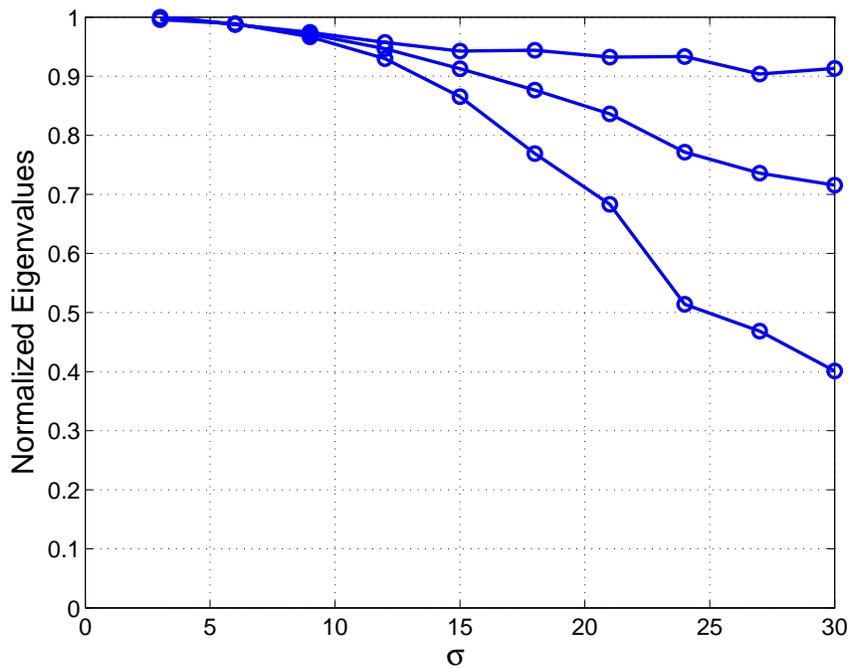}
%
% need bigger figs.  REFEREE MAY ASK FOR SOME INDICATION OF VARIABILIES OF NORM Evals
%
%\includegraphics[width=0.8\textwidth]{figures/RMS_vs_sigma.eps}
%\includegraphics[width=0.9\textwidth]{figures/sdpsvdoutlier.eps}
\caption{Plot of the eigenvalues of the OLS estimator $\hat S$ normalized by the eigenvalues of $S$ v.s. $\sigma$. Increasing the noise level biases the eigenvalues towards zero. A fragment of ubiquitin composed of 7 peptide planes (residue 1-8) and a specific Saupe tensor $S$ is used for the simulation and each point in the plot is computed from 200 different realizations of $\alpha_i,\beta_i$'s. ]
%{\bf Right:} Plot of $\langle \text{RMS}(r) \rangle_{\alpha_i,\beta_i}$ v.s. $\sigma$. The figure shows that the RMS of OLS residual has a near linear relation with the noise magnitude $\sigma$, when noise is added to the bond directions via torsion angle $\phi_i,\psi_i$'s.
}\label{fig:RMS_vs_sigma}
\end{figure}

\section{Numerical results}
\begin{figure}
\centering
\includegraphics[width=0.95\textwidth]{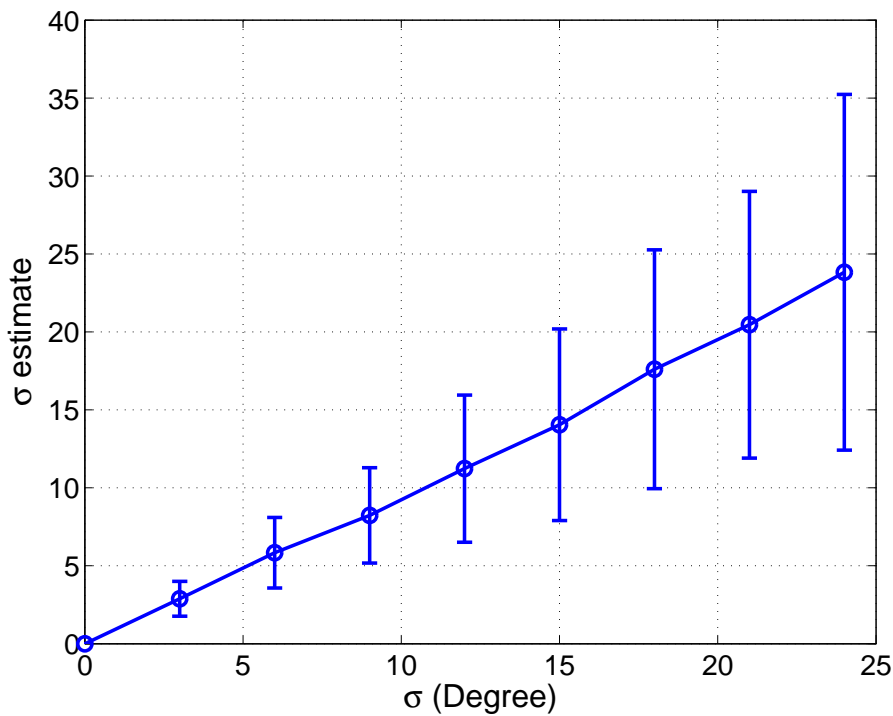}
\includegraphics[width=0.95\textwidth]{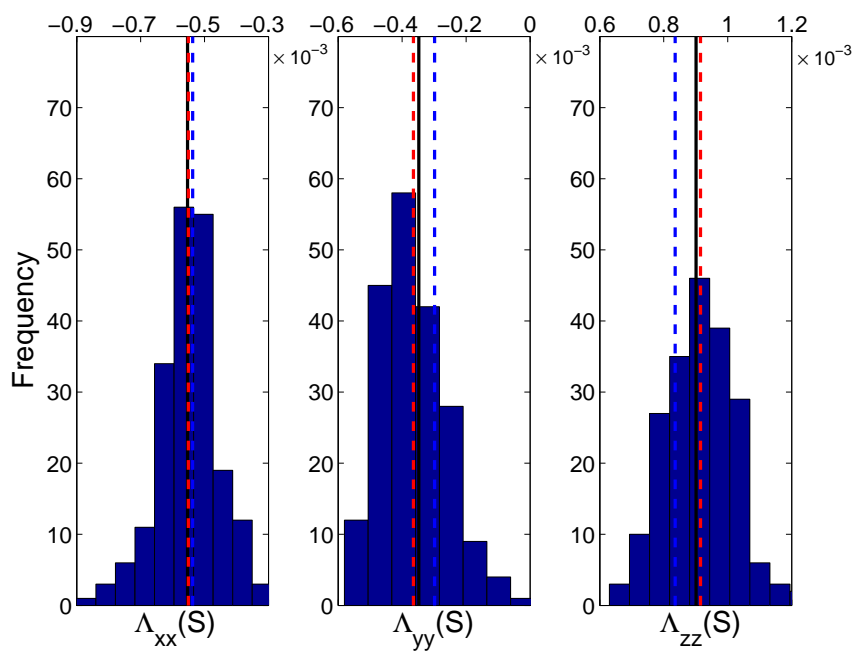}
\caption{{\bf Above:} Plot of $\hat \sigma$ v.s. $\sigma$. For a given noise level $\sigma$, $\hat \sigma$ is averaged over 200 different realizations of $\alpha_i,\beta_i$'s.
{\bf Below:} Histograms of the diagonal entries of $\Lambda(\hat S)$ and $\tilde \Lambda$ obtained from 200 fragment conformations with $20^\circ$ noise on the torsion angles. The values of $\Lambda(S), \langle \Lambda(\hat S)\rangle_{\alpha_i,\beta_i}$ and $\langle \tilde \Lambda \rangle_{\alpha_i,\beta_i}$ are denoted by black, blue and red line respectively.}\label{fig:sigma estimate}
\end{figure}

We first demonstrate that $\hat \sigma$ obtained through the method described in section \ref{section:sigma estimate} is a good estimate of $\sigma$. For simulation purposes, we use a segment of ubiquitin with seven peptide planes (residue 1-8) containing 21 $N-H$, $C-CA$ and $C-N$ bonds.
%We note that in all the simulations, we do not consider the RDC of $N-H$ bond for proline.
We note that in all the simulations, we do not consider the RDC of the nonexistent $N-H$ bond for proline.
In Fig.~\ref{fig:sigma estimate}(left), we plot $\hat \sigma$ v.s. $\sigma$. The simulation shows a close agreement between $\hat \sigma$ and $\sigma$, especially when the angular noise is less than 12 degrees.

We next show that the SIMEX-like method proposed in section \ref{section:method} is able to reduce the bias in eigenvalue estimation, where the bias of an estimator $\hat \theta$ of parameter $\theta$ is defined to be
$$\text{Bias}(\hat \theta) = \langle \hat \theta \rangle - \theta.$$
$\langle \cdot \rangle$ denotes averaging over the distribution of data. For this simulation, we use a specific ground truth Saupe tensor and the aforementioned ubiquitin fragment to generate precise RDC measurements. From the fragment, 200 realizations of noisy conformation are generated with $\sigma=20^\circ$. To obtain $\tilde \Lambda$, we set $n_1 = 8000$ when simulating $A_b$ in step (2) of the Monte Carlo procedure. In Fig.~\ref{fig:sigma estimate}(right), we see that the values of $\langle \tilde \Lambda \rangle_{\alpha_i,\beta_i}$ (Red dotted line) obtained from averaging over 200 samples are almost the same as the eigenvalues of $S$ (Black line), while there is a clear bias in the estimator $\Lambda(\hat S)$ (Blue dotted line).
%for different noise level $\sigma$. In The average of $\tilde \Lambda$

\subsection{Estimation of Saupe tensor eigenvalues from multiple molecular fragments}
While the proposed eigenvalue estimator $\tilde \Lambda$ has less bias, this does not obligate that $\tilde \Lambda$ has a lower mean squared error (MSE).This can be understood from the \emph{bias-variance decomposition}, which is a classical way in statistics to decompose the MSE of an estimator $\hat \theta$ \cite{wasserman2013all}.
 %{Manning, 2008 #4584}  	Manning CD, Raghavan P, Schütze H. Introduction to information retrieval: Cambridge university press Cambridge; 2008.
The MSE of an estimator $\hat \theta$ admits the following decomposition
\begin{eqnarray}
\text{MSE}(\hat \theta) &=& \langle (\theta - \hat \theta)^2 \rangle \cr
&=& \langle (\theta - \langle  \hat \theta \rangle + \langle  \hat \theta \rangle -\hat \theta)^2 \rangle\cr
&=& \text{Bias}(\hat \theta)^2 + \text{Var}(\hat \theta) + 2(\theta - \langle  \hat \theta \rangle )\langle \langle \hat \theta \rangle -\hat \theta \rangle\cr
&=& \text{Bias}(\hat \theta)^2 + \text{Var}(\hat \theta)
\end{eqnarray}
$\text{Var}(\hat \theta)$ denotes the variance of $\hat \theta$. Although we achieve less bias with the estimator $\tilde \Lambda$, we pay the price of having larger variance due to bias estimation involved in obtaining $\tilde \Lambda$. This increase in variance can lead to $\tilde \Lambda$ having higher MSE than $\Lambda(\hat S)$. From this point of view, when estimating the Saupe tensor eigenvalue using a single template fragment, the Monte Carlo method for debiasing may seem unnecessary or even disadvantageous. However, when multiple template fragments are available, the average of $\tilde \Lambda$ over these fragments, denoted $\tilde \Lambda_{\text{ave}}$, enjoys variance reduction proportional to the number of fragments. Therefore in the case when there are many fragments, it is worth paying the price of increased variance because the systematic bias error cannot be reduced via averaging. In the rest of the section, we use $\Lambda_{\text{ave}}(\hat S)$ to denote the average of $\Lambda(\hat S)$ over multiple fragments.

We now demonstrate the usefulness of our method under the setting of Molecular Fragment Replacement (MFR) approach \cite{delaglio2000mfr,kontaxis2005molecular}. When RDCs are measured in two different alignment medias for a protein of unknown structure, the MFR method can construct its structure by combining short homologous fragments obtained from chemical shift and dipolar homology database mining. Typically for every protein fragment of seven residues, 10 homologous structures are searched based on the similarity of chemical shifts and the goodness of Saupe tensor fit to the observed RDC. When OLS is used to fit the Saupe tensor with design matrix $A$ constructed from homologous structures, one can average all OLS eigenvalue estimated to obtain improved estimators of the parameters such as alignment magnitude and rhombicity that depend on the eigenvalues \cite{kontaxis2005molecular}. These parameters can in turn be used in a simulated annealing procedure such as XPLOR-NIH \cite{schwieters2003xplor} to refine the structure.

\begin{figure}
\centering
\includegraphics[width=0.9\textwidth]{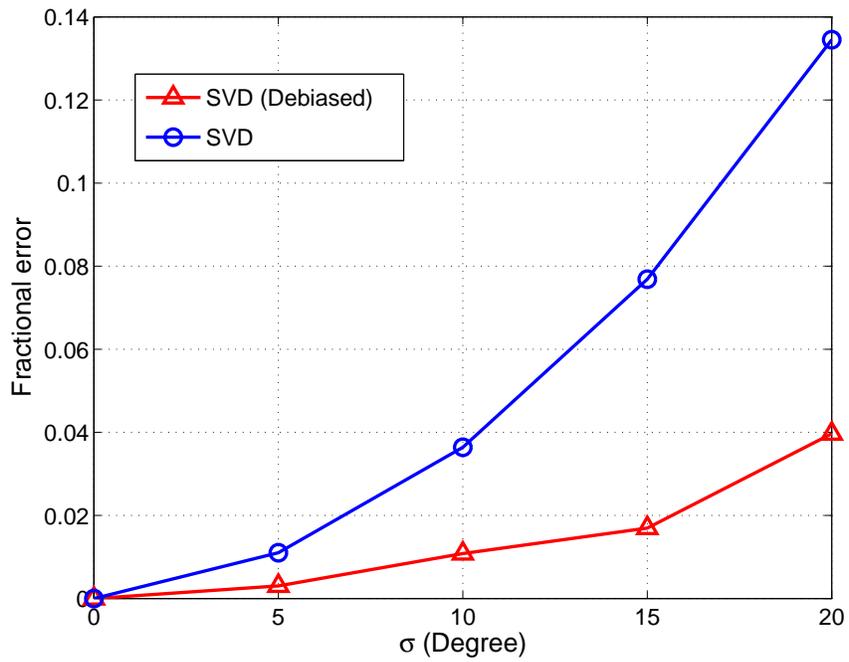}
\caption{Plot of the fractional error of $\Lambda_{\text{ave}}(\hat S)$ and $\tilde \Lambda_{\text{ave}}$ v.s. $\sigma$. Each data point is averaged over 12 different Saupe tensor and noise realizations for 1UBQ. The plot shows a clear advantage of the bias reduced estimator over the OLS estimator.}
\label{fig:MFR synthetic}
\end{figure}

We first use synthetic data to demonstrate our method. We generate 12 random Saupe tensors, by sampling two eigenvalues from the uniform distribution on $[-10^{-3},0]$ and $[0,10^{-3}]$ respectively, and extract the third eigenvalue by requiring $\Lambda(S)_{xx}+\Lambda(S)_{yy}+\Lambda(S)_{zz}=0$. The orthogonal matrix $U(S)$ is sampled uniformly from the group of $3\times 3$ orthogonal matrices, by computing the orthogonal factor in the polar decomposition of a $3 \times 3$ Gaussian random matrix \cite{blower2009random}. After obtaining the RDC $d_{nm}$'s from the clean structure and the ground truth Saupe tensor, under each simulated alignment condition we add structural noise of magnitude $\sigma$ to every fragment of seven peptide planes of the ubiquitin structure obtained from X-ray crystallography (PDB ID 1UBQ). We only consider the first 71 residues of the 76 residues of ubiquitin, as there are few RDC reported for the last five residues of ubiquitin. This gives a total of 64 fragments. We evaluate the estimators of the Saupe tensor eigenvalues $\Lambda_{\text{ave}}(\hat S)$ and $\tilde \Lambda_{\text{ave}}$ computed from the average of $\Lambda(\hat S)$ and $\tilde \Lambda$ of all fragments, by comparing their fractional errors averaged over the 12 different Saupe tensors and torsion angle noise realizations in Fig. \ref{fig:MFR synthetic}. The fractional error is defined as
$$\frac{\|\Lambda_{\text{ave}}(\hat S) - \Lambda(S)\|_F}{\|\Lambda(S)\|_F}\quad \text{and}\quad \frac{\|\tilde \Lambda_{\text{ave}} - \Lambda(S)\|_F}{\|\Lambda(S)\|_F}.$$
In this simulation, the fractional error of $\Lambda_{\text{ave}}(\hat S)$ is at least three times larger than $\tilde \Lambda_{\text{ave}}$.

\begin{figure}
\centering
\includegraphics[width=0.95\textwidth]{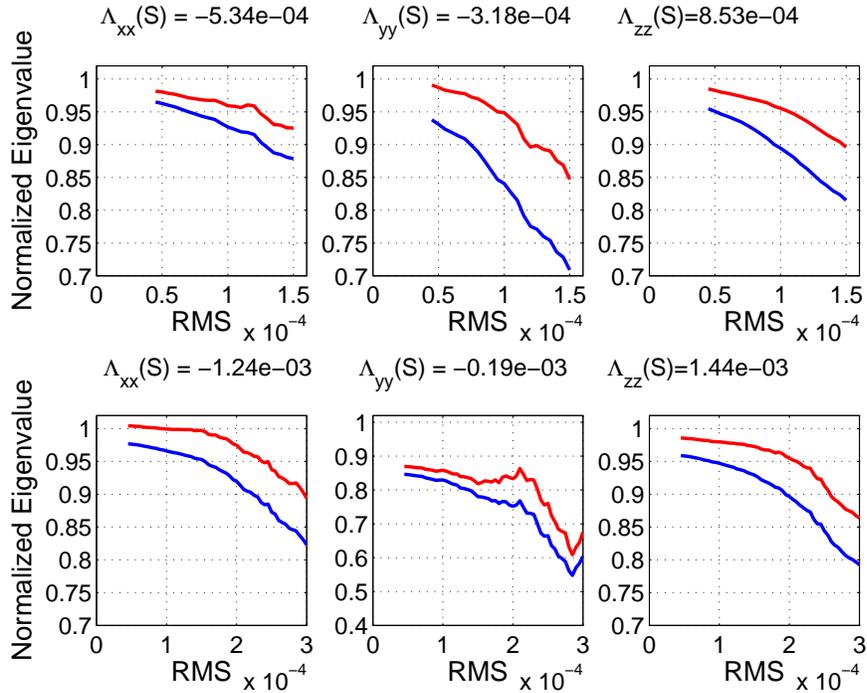}
\caption{Plot of the eigenvalue estimators normalized by $\Lambda(S)$ v.s. residual RMS thresholds. Estimators are obtained from experimental RDC measurements in two different alignment medias. While the magnitude of $\tilde \Lambda_{\text{ave}}$ (Red curves) and $\Lambda_{\text{ave}}(\hat S)$ (Blue dotted curves) both decrease as low quality (high RMS) fragments are solely used in averaging, $\tilde \Lambda_{\text{ave}}$ in general is within 90\% of the ground truth value but $\Lambda(S)$ drops to 80\% of $\Lambda_{\text{ave}}(\hat S)$. The value of $\Lambda(S)$ for both alignment medias are indicated in the plot title.}
\label{fig:MFR real}
\end{figure}

We finally apply this method to estimate the Saupe tensor of ubiquitin in two different alignment medias using the experimental RDC data in \cite{cornilescu19981D3Z}. From 600 homologous structures returned by MFR homology search, each containing seven residues, we obtain 600 Saupe tensor estimates using OLS. Since we expect our method to have a significant effect for fragments severely corrupted by structural noise, we average the fragments with residual RMS \emph{above} a certain threshold and plot $\Lambda_{\text{ave}}(\hat S)$ and $\tilde \Lambda_{\text{ave}}$ normalized by $\Lambda(S)$ v.s. RMS thresholds. To get an estimated (and approximate) ground truth Saupe tensor $S$, we use the high resolution ubiquitin structure 1UBQ obtained from X-ray crystallography \cite{vijay19871UBQ} to fit the RDC data. We demonstrate the results in Fig. \ref{fig:MFR real}. Other than the estimators for $\Lambda(S)_{yy}$ of the second alignment media which has a large percent error due to the relatively small magnitude of $\Lambda(S)_{yy}$, $\tilde \Lambda_{\text{ave}}$ typically achieves 0.9 of the ground truth value, whereas $\Lambda_{\text{ave}}(\hat S)$ can shrink to 0.8 of the value of $\Lambda(S)$ when only the fragments of high RMS are used in averaging. We therefore recommend the use of our proposed bias removing method when estimating eigenvalues from multiple noisy fragments.
%The comparison of the error in Tab. \ref{tab:MFR real}, in terms of $\|\Lambda_{\text{ave}}(\hat S) - \Lambda(S)\|_F$ and $\|\tilde \Lambda_{\text{ave}} - \Lambda(S)\|_F$,  for $\Lambda_{\text{ave}}(\hat S)$ and $\tilde \Lambda_{\text{ave}}$ obtained from averaging the eigenvalue estimates of all 600 fragments shows that the Monte Carlo method can achieve error reduction by more than a half.

\section{Effect of additive noise on Saupe tensor estimation}
\label{section:add noise}
So far we have been neglecting the presence of additive noise on $d_{nm}$, which is considered by \cite{losonczi1999order}. We define the noisy RDC measurements corrupted by additive noise as
\begin{equation}
\label{additive noise}
d_\text{add} = d + \sigma_\text{add} \varepsilon
\end{equation}
where entries of the column vector $\varepsilon$ are i.i.d random variables with mean zero. In this section, we show using perturbation theory that this type of additive noise biases the eigenvalue magnitude positively, therefore it cannot explain the magnitude shrinkage we see when fitting the Saupe tensor to real RDC data (Fig. \ref{fig:MFR real}). Moreover, the order of magnitude of this positive bias is not sufficient to explain the error between $\hat S$ and $S$. This has been noted by the authors of \cite{losonczi1999order} that in order to account for the size of the OLS misfit, an uncertainly of 2-3 Hz for the RDC measurements is required although the experimental uncertainty is only about 0.2-0.5 Hz.  This is the reason why in this paper we focus on removing the bias that arises from structural noise.

Let
\begin{equation}
S = U(S) \Lambda(S) U(S)^T
\end{equation}
be the eigendecomposition of $S$. Assuming the eigenvalues of $S$ are nondegenerate, the second order perturbation theory \cite{landau2013quantum} states
\begin{eqnarray}
\lambda_j(\hat S) &\approx&   \lambda_j(\Sgt) + U(\Sgt)_j^T (\hat S - S)  U(\Sgt)_j \cr
&\ & + \sum_{\substack{k=x,y,z,\\ k\neq j}} \frac{(U(\Sgt)_k^T  (\hat S - S) U(\Sgt)_j)^2}{\lambda_{j}(\Sgt)-\lambda_{k}(\Sgt)}.
\end{eqnarray}
Averaging the perturbation expansion over the distribution of $\varepsilon$, we get
\begin{eqnarray}
\label{bias PT}
\lambda_j(\hat S) \approx \lambda_j(\Sgt) + \sum_{\substack{k=x,y,z,\\ k\neq j}} \frac{\langle(U(\Sgt)_k^T  (\hat S - S) U(\Sgt)_j)^2\rangle_\varepsilon}{\lambda_{j}(\Sgt)-\lambda_{k}(\Sgt)}.
\end{eqnarray}
Here we use the fact that $\langle \hat S - S \rangle_\varepsilon = 0$ since
$$\langle \hat s - s \rangle_\varepsilon= \langle(A^TA)^{-1} A^T (A s + \varepsilon) - s \rangle_\varepsilon = \langle (A^TA)^{-1} A^T\varepsilon\rangle_\varepsilon = 0$$
The expression in (\ref{bias PT}) reveals that in the presence of noise, the largest eigenvalue of $\hat S$ is always greater than the largest eigenvalue of $\Sgt$, while the smallest eigenvalue behaves in the exact opposite manner. Such effect of bias of pushing the extreme eigenvalues outwards is also commonly seen in the context of estimating the extreme eigenvalues of covariance matrices \cite{schafer2005shrinkage}.

For this type of bias we now give an estimate of its order of magnitude. First we bound the numerator in the second order correction term in (\ref{bias PT}):
\begin{eqnarray}
\label{numerator bound}
(U(\Sgt)_k^T  (\hat S - S) U(\Sgt)_j)^2 &=& \Tr((\hat S - S)U(\Sgt)_j U(\Sgt)_k^T)^2\cr
&\leq& \|\hat S - S\|_F^2 \|U(\Sgt)_j U(\Sgt)_k^T\|_F^2\cr
&\leq& 3\|\hat s - s\|_2^2.
\end{eqnarray}
The first inequality results from Cauchy-Schwarz inequality, and the second inequality relies on the fact that $\|U(\Sgt)_j U(\Sgt)_k^T\|_F = 1$ and $\|\hat S - S\|_F^2 \leq 3\|\hat s - s\|_2^2$, which can be verified easily.
%$$\sum_{ij}(\hat S-S)_ij^2 = \|\hat s - s\|_F^2 +
It is a classical result \cite{gross2003linear} that the OLS estimator has covariance matrix
\begin{equation}
\langle (\hat s - s)(\hat s - s)^T \rangle_\varepsilon=  \sigma_\text{add}^2 (A^T A)^{-1},
\end{equation}
therefore
\begin{equation}
\label{s variance}
\langle \|\hat s - s\|_2^2 \rangle_\varepsilon= \sigma_\text{add}^2 \Tr((A^T A)^{-1}).
\end{equation}

Using (\ref{s variance}), (\ref{numerator bound}) we obtain an upperbound for the bias in (\ref{bias PT}). Taking $\lambda_z(\hat S)$ for example:
\begin{equation}
\label{actual estimate}
\lambda_z(\hat S) -  \lambda_z(\Sgt) \lesssim \frac{3\Tr((A^T A)^{-1})}{\lambda_z(S)-\lambda_y(S)} \sigma_\text{add}^2
\end{equation}
We now give an estimate of the order of magnitude of the bias. Since the magnitude of the extreme eigenvalues of the Saupe tensor is around $10^{-3}$, for example for the two RDC datasets acquired for ubiquitin, we simply assume $\lambda_z(S)-\lambda_y(S) \sim 10^{-4}$. The typical experimental uncertainty for RDC measurements is about 0.2 Hz - 0.5 Hz, and the dipolar coupling constant $D_{nm}^\text{max}$ for e.g. $N-H$ bonds, is about 23 kHz, therefore the noise magnitude $\sigma_\text{add}$ of the additive noise on the normalized dipolar coupling is about $0.5/(23 \times 10^3) \approx two \times 10^{-5}$. For ubiquitin, the average value of $\Tr((A^T A)^{-1})$ for fragments containing seven peptide planes is about 1.35. Using these numbers in  (\ref{actual estimate}), we get
$$ \lambda_z(\hat S) -  \lambda_z(\Sgt) \lesssim 1.6\times10^{-5},$$
which amounts to 1-2\% error when $\lambda_z(S)\sim10^{-3}$. This cannot explain the 10\% or larger error in fitting Saupe tensor to real RDC datasets using homology fragments in the previous section.

\begin{figure}
\centering
\includegraphics[width=0.95\textwidth]{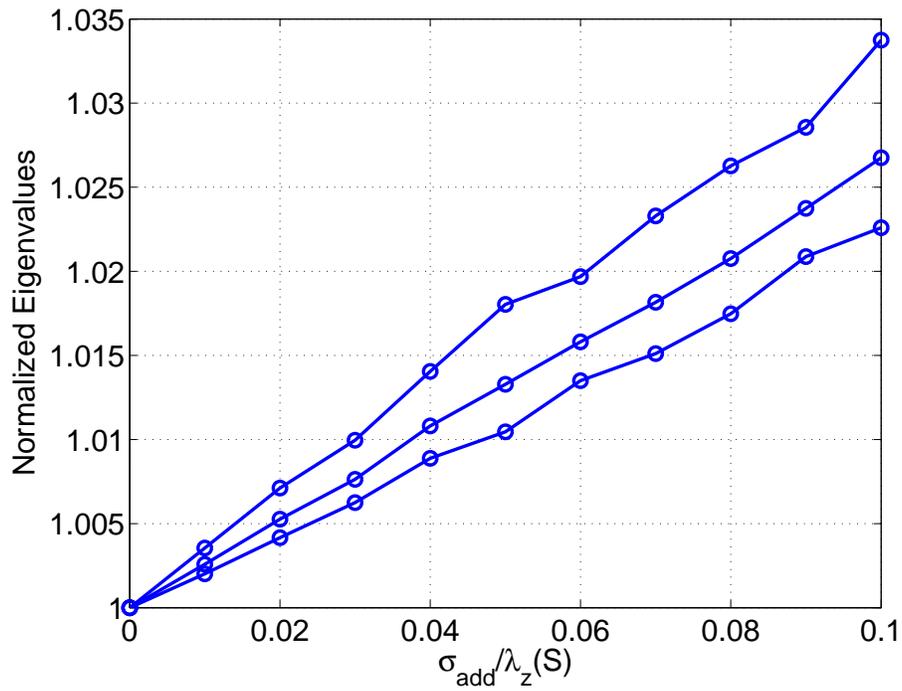}
\caption{Plot of the three eigenvalues of the OLS estimator $\hat S$ normalized by the eigenvalues of $S$ v.s. $\sigma_\text{add}$ under noise model (\ref{additive noise}). Each point is averaged over 2000 noise and Saupe tensor realizations. Increasing the noise level biases the eigenvalues positively, unlike the case for structural noise. At 10\% noise level, the bias is about 3\%.}\label{fig:Bias add}
\end{figure}

We present a simulation to illustrate the bias in OLS eigenvalues estimation in the presence of additive noise. We use the Saupe tensor eigenvalues for ubiquitin in the first alignment media presented in Fig.~\ref{fig:MFR real}, and a ubiquitin fragment consisting of seven peptide planes (residue 1-8) for this simulation. We generate noisy datasets using the noise model
$$d_\text{add} = As + \sigma_\text{add} \varepsilon.$$
For every noise level, we average $\Lambda(\hat S)$ normalized by $\Lambda(S)$ over 500 different realizations of $s$ and $\varepsilon$ where entries of $\varepsilon$ are i.i.d. random normal variables. The different realization of $s$ are generated from $S = U(S) \Lambda(S) U(S)^T$ where $\Lambda(S)$ is fixed but $U(S)$ is sampled uniformly from the orthogonal group in $\mathbb{R}^3$. We vary the orientation of the Saupe tensor since it is clear from (\ref{bias PT}) that the bias depends on $U(S)$. We change $\sigma_\text{add}$ from 0 to 10\% of $\lambda_z(S)$ and present the results in Fig. \ref{fig:Bias add}. We note again from previous calculations, $\sigma_\text{add}\sim  2 \times 10^{-5}$, which amounts to 2-3\% of the $\lambda_z(S) = 0.85\times 10^{-3}$ considered. As shown in the simulation and our crude estimate, such magnitude of noise gives rise to bias error of about 1\%. Even in the case of having very noisy RDC (having noise magnitude 10\% of $\lambda_z(S)$), the bias error caused by additive noise is around 3\%. Whereas in a typical MFR search with torsion angle tolerance being set to $\pm 20^\circ-30^\circ$ \cite{wu2005mfr}, the simulation in Fig.~\ref{fig:RMS_vs_sigma} suggests structural noise can cause bias error sometimes much greater than 10\%. Therefore in this paper we focus on removing the bias that arises from structural noise. In the case when accurate template structure is available and the additive noise is a concern, we refer readers to the appendix for the removal of such bias using an analytic expression derived from perturbation theory.

%We note that it is simple to estimate the bias for the middle eigenvalue $\lambda_y(\hat S)$, as
%\begin{eqnarray}
%\langle \lambda_y(\hat S)\rangle_\varepsilon &=& \langle-\lambda_x(\hat S) + \lambda_z(\hat S)\rangle_\varepsilon \cr
%&=& -\lambda_x(S)+\lambda_z(S)-\text{Bias}(\lambda_x(\hat S))+\text{Bias}(\lambda_z(\hat S))\cr
%&=& \lambda_y(S)-\text{Bias}(\lambda_x(\hat S))+\text{Bias}(\lambda_z(\hat S)).
%\end{eqnarray}
%Due to the substraction, the bias for $\lambda_y(S)$ can be much less than $10^{-5}$.

%\begin{eqnarray}
%\langle \|\hat s - s\|_2^2 \rangle_\varepsilon &=& \langle \|(A^T A)^{-1} A^T d_\text{add} - s\|_2^2 \rangle_\varepsilon\cr
%&=& \langle \|(A^T A)^{-1} A^T (A s + \varepsilon) - s\|_2^2\rangle_\varepsilon\cr
%&=& \langle \| A^T \varepsilon)\|_2^2\rangle_\varepsilon\cr
%&=&  \Tr ( (A^T A)^{-1} A^T \langle \varepsilon \varepsilon^T \rangle_\varepsilon A (A^T A)^{-1})\cr
%&=& \sigma_\text{add}^2 \Tr((A^T A)^{-1})
%\end{eqnarray}

\section{Conclusion}
We observe a negative bias when estimating the Saupe tensor eigenvalues through the classical SVD method, in the presence of structural noise on the template structure due to torsion angle noise. We present a Monte Carlo method that simulates noise on the template structure by perturbing the torsion angles and use the simulated structure to estimate the bias in the eigenvalues. We demonstrate the effectiveness of our method in reducing the error arising from bias when estimating Saupe tensor eigenvalues from multiple protein fragments, which is a natural setting to consider when building protein structure from homologous substructures.

\section{Acknowledgement}
The research of AS was partially supported by award R01GM090200 from the NIGMS, by awards FA9550-12-1-0317 and FA9550-13-1-0076 from AFOSR, by award LTR DTD 06-05-2012 from the Simons Foundation, and the Moore Foundation Data Driven Discovery Investigator award.


\section{Appendix}
\subsection{Removing bias from additive noise}
Define a linear operator $L: \mathbb{R}^{5} \rightarrow \mathbb{R}^{3\times 3}$ that forms a Saupe tensor $S$ from the vector $s$ as
\begin{equation}
\label{linear operator L}
L(s) = \begin{bmatrix} -s(1)-s(2) & s(3) & s(4) \\ s(3) & s(1) & s(5) \\ s(4) & s(5) & s(2) \end{bmatrix},\quad s \in \mathbb{R}^5.
\end{equation}
For the additive noise model (\ref{additive noise}) we have
\begin{equation}
\label{S OLS}
\hat S = L(\hat s) = L ((A^T A)^{-1} A^T d_\text{add}) = S +  L ((A^T A)^{-1} A^T \varepsilon).
\end{equation}
We also define the adjoint operator of $L$, $L^*:\mathbb{R}^{3\times3} \rightarrow \mathbb{R}^5$ through the relation
\begin{equation}
\label{adjoint L}
\Tr(X^T L(y)) = L^*(X)^T y,
\end{equation}
for every $y\in \mathbb{R}^5$ and $X \in \mathbb{R}^{3\times 3}$. To obtain the form of $L^*$, we let $y\in \{e_1,\ldots,e_5\}$, where $e_i(i)=1$ and $e_i(j) = 0$ if $j\neq i$. Plugging such $y$ into Eq. (\ref{adjoint L}), we get
\begin{eqnarray}
{L^*(X)}_1 &=& -X_{xx}+X_{yy} \cr
{L^*(X)}_2 &=& -X_{xx}+X_{zz} \cr
{L^*(X)}_3 &=& X_{xy} + X_{yx} \cr
{L^*(X)}_4 &=& X_{xz} + X_{zx} \cr
{L^*(X)}_5 &=& X_{yz} + X_{zy}
\end{eqnarray}
Using such notion of the adjoint operator, the perturbation series in (\ref{bias PT}) can be written as
\begin{multline}
\label{PT Taylor}
\langle \lambda_j(\hat S) \rangle_\varepsilon \approx \lambda_j(\Sgt) +  \sum_{\substack{k=x,y,z,\\ k\neq j}}  \frac{ \langle ((\hat s - \sgt)^T L^*(U(\Sgt)_k U(\Sgt)_j^T))^2 \rangle_\varepsilon}{\lambda_{j}(\Sgt)-\lambda_{k}(\Sgt)}\\
=  \lambda_j(\Sgt) +   \Tr\bigg[ \bigg(\sum_{\substack{k=x,y,z,\\ k\neq j}}  \frac{L^*(U(\Sgt)_k U(\Sgt)_j^T)L^*(U(\Sgt)_k U(\Sgt)_j^T)^T}{\lambda_{j}(\Sgt)-\lambda_{k}(\Sgt)}\bigg) \text{Var}(\hat s)\bigg]
\end{multline}
where
\begin{equation}
\text{Var}(\hat s) \equiv \langle (\hat s - \sgt)(\hat s - \sgt)^T\rangle_\varepsilon = (A^TA)^{-1} \sigma_\text{add}^2.
\end{equation}
Therefore we can subtract the second order term in (\ref{PT Taylor}) to correct for the bias in the eigenvalues. Although we do not know the eigenvectors and eigenvalues of $\Sgt$, we can replace them with the eigenvectors and eigenvalues $\hat S$. This change will only affect on the higher order terms in the perturbation series.


%\bibliographystyle{elsarticle-num}
\bibliographystyle{amsplain}
\bibliography{bibref}


%\begin{thebibliography}{00}


%\bibitem{}

%\end{thebibliography}
\end{document}